\documentstyle[epsf]{l-aa}
\newcommand{\eqb}{\begin{eqnarray}}
\newcommand{\eqe}{\end{eqnarray}}
\newcommand{\sigmaT}{\sigma_{\rm T}}
\newcommand{\om}{\omega}
\newcommand{\al}{\alpha}
\newcommand{\th}{\Theta}
\newcommand{\md}{\mbox{d}}
\newcommand{\nn}{\nonumber}
\newcommand{\ul}{\underline}
\newcommand{\matr}[1]{\underline{\underline{\tens{#1}}}}
\newcommand{\vzw}{\langle v^2 \rangle}
\newcommand{\vvi}{\langle v^4 \rangle}
\newcommand{\dsty}{\displaystyle}
\newcommand{\melec}{m_{\rm e}}
\newcommand{\nelec}{n_{\rm e}}
\newcommand{\Telec}{T_{\rm e}}
\newcommand{\kB}{k_{\rm B}}
\newcommand{\sqy}{\sqrt{1-y^2}}
\newcommand{\Jpar}{J_{\parallel}}
\newcommand{\Jperp}{J_{\perp}}
\begin{document}
\thesaurus{12(02.18.5; 02.18.7; 03.13.1; 11.01.2; 13.25.2; 13.25.5)}
\title{An eigenfunction method for the comptonisation problem}
\subtitle{Angular distribution and spectral index 
of radiation from a disk}
\author{U.D.J. Gieseler \& J.G. Kirk}
\institute{Max-Planck-Institut f\"ur Kernphysik,
Postfach 10 39 80, D-69029 Heidelberg, Germany}
\offprints{U.D.J. Gieseler}
\date{Received \dots}
\maketitle
\begin{abstract}  
We present a semi-analytic approach to solving the Boltzmann equation 
describing the comptonisation of
low frequency input photons by a thermal distribution of electrons 
in the Thomson limit. 
Our work is based on the formulation of the problem by 
Titarchuk \& Lyubarskij~(\cite{TitarLyu95}), but extends their treatment 
by accommodating an arbitrary anisotropy of the source function. 
To achieve this, we expand the eigenfunctions of
the integro/differential eigenvalue problem defining the spectral  
index of comptonised radiation in terms
of Legendre polynomials 
and Chebyshev polynomials. 
The resulting algebraic 
eigenvalue problem is then solved by numerical means, yielding 
the spectral index and the full angular and spatial 
dependence of the specific intensity of radiation.
For a thin ($\tau_0 < 1$) plasma disk, the radiation is 
strongly collimated along the 
disk surface -- for an optical thickness of $\tau_0 =0.05$, the radiation 
intensity along the surface is roughly ten times that along the 
direction of the normal, and varies only slightly  with the electron 
temperature. Our results for the spectral index confirm those of 
Titarchuk \& Lyubarskij over a wide range of electron temperature and 
optical depth; the largest difference we find is roughly 
10\% and occurs at low optical depth.

\end{abstract}

\section{Introduction}
\label{intro}
The process of Comptonisation, in which soft photons increase their 
energy by scattering in a gas of hot electrons is thought to be the 
main mechanism responsible for the formation of non-thermal continuum 
spectrum in a range of astrophysical objects. Much of the 
early work on this subject assumed that the optical depth of the 
scattering cloud was fairly large, and that the photon frequency 
 $\nu$ and electron temperature $\Telec$ were both small: $x\equiv
 h\nu/\melec c^2\ll 1$, \mbox{$\th\equiv\kB\Telec/\melec c^2\ll 1$}. The 
transport of a photon in both configuration space and in energy space 
can then be approximated by a Fokker-Planck equation and the computation 
of the spectrum can be split into two parts -- the evaluation of the 
distribution of the number of scatterings undergone by a photon, and the 
convolution of this with the solution for the evolution of the spectrum 
in an infinite homogeneous medium (Sunyaev \& Titarchuk \cite{SunTitar80}).   

In the wake of this important result, many authors have investigated 
methods of extending the range of applicability of the calculations, 
e.g., by calculating corrections to the diffusion coefficient in energy 
space to higher order in the parameters $\th$ and $x$ (Prasad et 
al.~\cite{Prasad88}; see also Cooper~\cite{Cooper71}). Of 
particular interest is the generalisation to scattering media of 
optical depth $\tau\sim 1$, since such conditions are indicated in many 
applications (e.g., AGN: 
Haardt et al.~\cite{haardtetal94}; Zdziarski et al.~\cite{zdziarskietal95}
and galactic black hole candidates 
Sunyaev \& Tr\"umper \cite{SunTruem79}; Ebisawa et al.~\cite{ebisawaetal96}).
Thus, Sunyaev \& 
Titarchuk (\cite{SunTitar85}) presented a numerical solution to 
the spatial transport of photons in slab geometry. Assuming the problem 
remains separable, this can be combined with the solution for diffusion 
in energy space to yield the emergent spectrum. These and other 
generalisations are summarised by Titarchuk (\cite{Titar94}), who 
also pointed out that the angular distribution of comptonised radiation 
emerging from an optically thin disk forms a \lq knife-blade\rq\ 
pattern, collimated almost parallel to the surface of the disk. In 
addition to analytic work, comptonisation has been investigated
using numerical methods (Katz \cite{Katz76}; 
Poutanen \& Svensson~\cite{poutanensvensson96}), in particular
the Monte-Carlo simulation technique (Pozdnyakov et al.~\cite{Pozdny83}; 
Zdziarski \cite{Zdzia86}; Hua \& Titarchuk \cite{HuaTitar95}; 
Stern et al.~\cite{sternetal95a,sternetal95b}). 

In a recent paper, Titarchuk \& Lyubarskij (\cite{TitarLyu95}) have attacked 
the problem 
using the Boltzmann equation, without recourse to a Fokker-Planck 
approximation in either configuration or energy space. As well as 
confirming the result that power-law spectra are produced when low 
frequency photons are scattered in the Thomson regime (i.e, when the 
dimensionless photon energy $x'$ measured in the electron rest frame 
satisfies $x'\ll1$), they demonstrate explicitly the separation of the 
problem into its configuration and energy space parts, provided that 
the {\em source function} can be considered isotropic. In this case,
the problem reduces to an eigenvalue equation for the source
function. Titarchuk \& Lyubarskij also provide 
analytic approximations for the computation of the 
power-law index $\al$ over a very large range of optical depth and 
temperature.  

In this paper we generalise the approach of Titarchuk \& 
Lyubarskij (\cite{TitarLyu95}) by solving the Boltzmann 
equation without assuming isotropy of either the radiation or the 
source function. The approach we adopt is to formulate the equation 
determining the power-law index as an integral eigenvalue problem. To 
find the eigenfunctions, we expand the angular dependence in a series 
of Legendre polynomials, and the spatial dependence in a series of 
Chebyshev polynomials. Our results confirm the accuracy of the formulae 
presented by Titarchuk \& Lyubarskij (\cite{TitarLyu95}) and 
also give the full spatial and angular dependences of the comptonised 
radiation. As predicted by Titarchuk (\cite{Titar94}), the 
radiation from an optically thin slab is strongly collimated along the 
surface of the slab.
                      
The paper is laid out as  follows: in Sect.~\ref{formu} we present the
equations leading to the formulation of the integral eigenvalue problem for 
the power-law index $\al$. Section \ref{method} (and in more detail the 
appendix) presents the method of solution. 
The phase function is first of all expanded in terms of Legendre polynomials 
of the scattering angle and the integration over electron velocity 
is performed in the case of low temperature
($\th\ll1$) and high temperature ($\th\gg 1$).
Then, using an expansion in Legendre polynomials which are complete over the
 {\rm half-range} $0 \le \mu \le 1$ of the cosine of the angle 
between the photon direction and the normal to the slab, the problem is 
converted to a system of differential equations in the spatial 
coordinate normal to the disk surface. 
This is finally reduced to an algebraic eigenvalue 
problem by 
expanding the spatial dependence in a series of Chebyshev polynomials. 
Singular value decomposition is then used to find the eigenvalues and 
eigenfunctions. Our results are presented in Sect.~\ref{results}. 
These consist of 
plots of the spectral index as a function of temperature $\th$ and 
half-thickness $\tau_0$ of the slab, the angular dependence of the 
specific intensity of radiation, and its spatial distribution. We 
compare these to the formulae for $\al$ presented by Titarchuk \& 
Lyubarskij (\cite{TitarLyu95}).
Section \ref{discussion} contains a summary of our conclusions, and a short 
discussion of their range of applicability in astrophysical sources. 

\section{Formulation of the basic equations}
\label{formu}
Let us first define the geometry of the problem. We consider an infinite 
disk of thickness $2z_0$ containing a uniform, non-degenerate gas of 
free-electrons of temperature 
$\Telec$ and number density $\nelec$. The only process of importance 
for the transport of photons in this disk is Compton scattering. The 
optical half-thickness of the disk is defined as $\tau_0=\sigmaT\nelec z_0$, 
where $\sigmaT=6.65\times 10^{-25}\,{\rm cm}^2$ is the Thomson 
cross-section. Let the spatial coordinate normal to the disk 
be $z$, with $z=0$ on the mid-plane of the disk, and define the optical 
depth variable $\tau=\nelec\sigmaT z$ which is bounded by
 $-\tau_0\le\tau\le\tau_0$.

We denote the cosine of the angle between the normal to the disk
and the direction of a photon after scattering by $\mu$. In the same sense,
the direction before a scattering event is denoted by $\mu_1$.
Let $\eta$ be the cosine of the angle between these directions. 
For an isotropic electron distribution, the phase function,
which describes the change in photon direction due to scattering, depends
only on $\eta$.
To calculate this phase function, we have to perform an integral over the
electron velocity. Because of the isotropy of the electron distribution,
the electron direction does not refer to the disk normal.
It is therefore convenient to switch to a coordinate system, in which the 
electron direction defines the $z$-axis. 
The cosines of the photon directions with respect to the 
electron direction are then denoted by $\tilde\mu$ (after scattering), and
 $\tilde\mu_1$ (before scattering). An analogous notation is used for 
the azimuthal angles.

We are interested in a situation in which low 
frequency radiation is injected into the 
disk, is scattered by the electrons, and forms a power-law spectrum at 
high frequency, as shown 
by Shapiro et al.~(\cite{Shapiro76}) and Sunyaev \& Titarchuk
(\cite{SunTitar80}), and seen in numerical studies of Katz 
(\cite{Katz76}). In the power-law regime, there is no source of 
radiation in the disk, and no radiation which enters from the outside.
The time independent, polarisation averaged, equation of transfer for the 
specific (up-scattered)
intensity $I(\nu,\mu,\tau)$ is then given by 
\eqb \label{transfer}
\lefteqn{
\mu\,\frac{\partial}{\partial \tau}I(\nu,\mu,\tau) =
       \frac{1}{n_{\rm e}\sigmaT}
    \int\limits_{0}^{\infty}\md\nu_1\int\limits_{4\pi}\md\Omega_1 } \nn\\
&&\Bigl[\frac{\nu}{\nu_1} \sigma_{\rm s}(\nu_1\rightarrow\nu,\eta)
     I(\nu_1,\mu_1,\tau) - \sigma_{\rm s}(\nu\rightarrow\nu_1,\eta)
     I(\nu,\mu,\tau)\Bigr]\nn\\&&
\eqe
(see e.g., Pomraning \cite{Pomraning73}).
This transport equation is a special case of the linearised Boltzmann 
equation and we shall simply refer to it as the Boltzmann equation.

The power-law part of the spectrum we describe occurs at frequencies 
lower than that of the Wien cut-off ($h\nu < \kB \Telec$), so that the
energy change of the photon due to the recoil of the electron, can be neglected
in comparison with the Doppler shift of the photon.
In the electron rest frame (in which quantities are adorned with a prime), the 
classical Thomson scattering kernel is then given by
\eqb \label{sigma}
\sigma_{\rm s}'(\nu'\rightarrow\nu_1',\eta')={3\over 16 \pi}n'_{\rm e}\sigmaT
\left\lbrace 1+(\eta')^2\right\rbrace \,\delta(\nu'-\nu_1')\,.
\eqe
The electrons are distributed isotropically. They are described by a 
relativistic Maxwell distribution, which is defined as
\eqb \label{maxwell}
f(v)=\frac{\gamma^5\,\exp(-\gamma/\th)}
          {4\pi\,\th\,{\rm K}_2(1/\th)}\,,
\eqe
where ${\rm K}_2$ is the modified Bessel function of the second kind of order
two, and $\gamma=(1-v^2)^{-1/2}\,,$ (here and in the following we set $c=1$).
This distribution is normalised according to
\eqb \label{f_norm}
\int\md^3 v\,f(v) = 4\pi \int\md v\,v^2\,f(v) = 1 \,.
\eqe

The scattering kernel in the disk system can be calculated by performing
a Lorentz boost of $\sigma_{\rm s}'$, multiplying it by $f(v)$ and integrating
over $v$. Then the scattering kernel is given by\footnote{See also
Titarchuk \& Lyubarskij (\cite{TitarLyu95}) Eq.~(2) after correction of a
minor typographical error.}
\eqb \label{sigma_disk}
\lefteqn{
\sigma_{\rm s}(\nu\rightarrow\nu_1,\eta,\th)
 \,=\, \frac{3}{16\pi}\frac{n_{\rm e}\sigmaT}{\nu x}\int\!\md^3 v\,
             \frac{f(v)}{\gamma}
         }  \nn \\
&&\quad\cdot\Bigl\{ 1+\Bigl(1\!-\!\frac{1-\eta}{\gamma^2 D D_1}\Bigr)^2\Bigr\} 
\cdot\delta\Bigl[\gamma\Bigl(\frac{D}{x_1}-\frac{D_1}{x}
                                                            \Bigr)\Bigr]\,,
\eqe
with the definitions 
\eqb \label{D_def}
 D   &:=& 1-\tilde\mu\,v\,,\nn\\
 D_1 &:=& 1-\tilde\mu_1\,v \,. 
\eqe
We now look for a solution of the Boltzmann equation of the form 
\eqb \label{ItoJ}
 \mbox{$I$}(\nu,\mu,\tau)  = J(\mu,\tau)\,x^{-\al} \,.
\eqe
Inserting this, and Eq.~(\ref{sigma_disk}), into Eq.~(\ref{transfer}), and 
performing the trivial integral over $\nu_1$ using the $\delta$-
function, we find
\eqb \label{Boltzmann}
\mu\frac{\partial J(\mu,\tau)}{\partial\tau} = -J(\mu,\tau) + B(\mu,\tau)\,,
\eqe
where the source function $B(\mu,\tau)$ is defined as
\eqb \label{source}
B(\mu,\tau) = \frac{1}{4\pi}\int\limits_{-1}^{1}\md\mu_1
              \int\limits_{0}^{2\pi}\md\phi \, R(\eta)\,J(\mu_1,\tau)\,.
\eqe
The phase function $R(\eta)$ is given by
\eqb \label{phase}
R(\eta)=\frac{3}{4}\int\md^3 v\,\frac{f(v)}{\gamma^2}
    \Bigl(\frac{D_1}{D}\Bigr)^{\al+2}\frac{1}{D_1}\bigl\{1+(\eta')^2\bigr\}\,. 
\eqe
The last three equations (\ref{Boltzmann}, \ref{source} and 
\ref{phase}), together with the boundary condition 
\eqb \label{boundary}
J(\mu < 0,\tau =\tau_0) \equiv\, 0\, \equiv J(\mu > 0, \tau = -\tau_0) \,,
\eqe
define an integral eigenvalue problem. Our aim 
is to reduce this to an algebraic eigenvalue equation by expanding the 
intensity $J(\mu,\tau)$ into a polynomial series, as mentioned above and shown
in the Sects.~\ref{angular_dep} and \ref{spatial_dep}.
We further perform the phase function integral using an expansion for
 $\th\ll 1$ and $\th \gg 1$. This is shown in the Sects.~\ref{theta_small} 
and \ref{theta_big}.
\section{Method of solution}
\label{method}
To reduce the integral eigenvalue problem described above to an 
algebraic eigenvalue equation, we first expand the phase 
function in a series of Legendre polynomials:  
\eqb \label{R_eta_1}
 R(\eta)  = \sum_{i=0}^M \om_i(\al,\th) P_i(\eta) \,.
\eqe
The details of this and the following steps are shown in the appendix.
The coefficients $\om_i(\al,\th)$ can be written, in the case of small
electron temperature ($\th \ll 1$), as polynomials in $\al$, 
with coefficients in terms of moments of the Maxwell distribution, which
can easily be calculated numerically.
In the case of high electron temperature ($\th \gg 1$), the coefficients
 $\om_i(\al,\th)$ can be expressed using the incomplete $\Gamma$-function.

The essential step is to expand the specific intensity $J(\mu,\tau)$ into a 
series of Legendre and Chebyshev polynomials. For the half-space
 $0\le\mu\le 1$, we write (compare Eq.~(\ref{J_expand})) 
\eqb \label{J_expand_1}
J(\mu,\tau)\big|_{\mu \ge 0} &=& \sum_{n=0}^N \frac{2 n+1}{2} P_n(2\mu-1)
                                        Q_n^+(\tau)  \,.
\eqe
We now expand each of these $N+1$ expansion coefficients 
 $Q^+_n(\tau)$ into 
a series of Chebyshev polynomials ($T_i$). In vector-form this reads
(compare Eq.~(\ref{Q_expand}))
\eqb \label{Q_expand_1}
\vec{Q}^+(\tau/\tau_0)=\vec{q}_0 +\sum_{i=1}^K 
                 \bigl[ 1-T_i(-\tau/\tau_0)\bigr]\,\vec{q}_i\,.
\eqe
All expansion coefficients can be represented by a common $(K+1)\cdot (N+1)$
dimensional vector: 
\eqb \label{q_vec_1}
\ul{\vec{q}} := \left( \begin{array}{c}
                \vec{q}_0 \\ \vec{q}_1 \\ \vdots \\ \vec{q}_K 
                 \end{array} \right) \,\, .  
\eqe
As shown in detail in the appendix,
the eigenvalue problem (Eqs.~\ref{Boltzmann}-\ref{phase}) becomes with
these expansions a set of homogeneous linear equations for the vector
 $\ul{\vec{q}}$, which is very easy to treat numerically. The set of equations
can be written as a matrix equation (see Eq.~(\ref{T_eq_short})): 
\eqb \label{T_eq_short_1}
\matr{F}(\al,\th,\tau_0)\,\ul{\vec{q}}\,=0\,. 
\eqe
 The first step is to calculate the matrix 
 $\matr{F}(\al,\th,\tau_0)$ for given values of $\tau_0$ and $\th$.
The solution for the spectral index $\al$ can be found by solving the following
equation for the determinant of $\matr{F}(\al)$\,:
\eqb \label{determinant}
\mbox{det}\bigl[\matr{F}(\al)\bigr]\stackrel{!}{=}0 \,.
\eqe
We used {\em Mathematica} to find the roots of this equation, which give the 
spectral index $\al$. For an expansion of the spatial and angular dependence of
 $J(\mu,\tau)$ to (e.g.) 9th order, $\ul{\vec{q}}$ is a 100 dimensional vector
and $\matr{F}(\al)$ is a $100\times 100$ matrix. A solution of 
Eq.~(\ref{determinant}) for this dimension takes less than one minute
with a {\em Pentium} PC in both the relativistic, and the 
non-relativistic cases.

For each value of $\al$, which satisfies Eq.~(\ref{determinant}), the expansion
coefficients, and therefore $J(\mu,\tau)$, can be found
by solving the equation
\eqb \label{Fq=0}
\matr{F}\,\,\ul{\vec{q}} =0 \,,
\eqe
where $\matr{F}$ is now a known singular square matrix. This matrix 
can be decomposed using a singular value decomposition routine (see e.g.
Press et al.~\cite{Press86}; Wolfram \cite{Wolfram91}), which yields 
the vector null-space.
This vector can immediately inserted into Eq.~(\ref{Q_expand_1}), which gives
 $\vec{Q}^{\pm}(\tau)$. Inserting this into Eq.~(\ref{J_expand_1}), we end up
with the intensity $J(\mu,\tau)$. The resulting values for the spectral
index $\al$ and the shape of the intensity in $\mu$ and $\tau$ are 
described in the next section. Use of the singular value decomposition 
technique has the advantage that an automatic check for the presence of 
repeated  roots of Eq.~(\ref{determinant}) is provided. More importantly, 
however, it immediately provides the 
null-space and the associated specific intensity of 
radiation, which must be physically realistic in the sense that $J(\mu,\tau)$ 
must be positive definite. This important condition 
enables one to identify the 
physically relevant power-law index $\alpha$, 
which is the smallest positive root 
of the nonlinear Eq.~(\ref{determinant}).
\section{Results}
\label{results}
\subsection{Spectral index}
\label{spectral}
\begin{figure}[t]
    \vspace{-2.7cm}
    \begin{center} 
      \epsfxsize9.0cm 
      \mbox{\epsffile{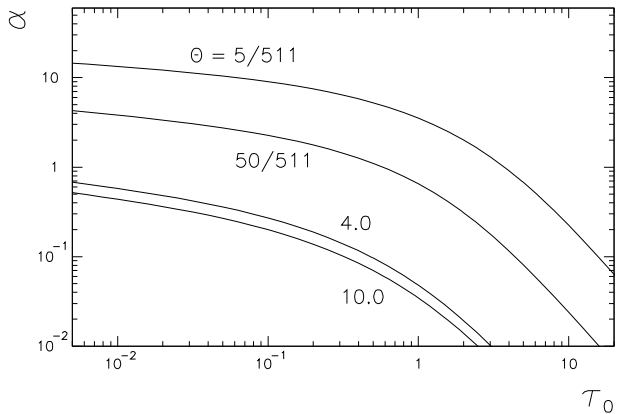}}
    \end{center}
    \vspace{-1cm} 
    \caption{Spectral index $\al$ vs. Thomson optical half thickness
             $\tau_0$, for non-relativistic and relativistic electron 
             plasma temperatures $\th =\kB\Telec/m_{\rm e}\approx
\kB\Telec/(511\,{\rm keV})$.}
    \label{al_tau_plot}
\end{figure}
The spectral index $\al$ of the comptonised radiation, for a given
temperature ($\th\ll 1$ or $\th\gg 1$), and {\em arbitrary} optical depth
$2\tau_0$, can be found by solving Eq.~(\ref{determinant}). 

Note that the spectral index is the exponent of the $v$-dependent functions
 $D$ and $D_1$ in the phase function. For non-relativistic plasma 
temperatures, the spectral index can be well above 1.
In this case, accuracy is achieved only if the phase
function is expanded to high order in a Taylor series in $v$ -- 
we use an 
expansion up to 16th.\ order  ($L=16$ in Eq.~(\ref{taylor})). This sum
is represented by an expansion to 10th.\ order in $\eta$ ($M=10$ in 
Eq.~(\ref{taylor})). In the relativistic temperature regime, on the other 
hand, the spectral index $\alpha$ is much lower. 
Therefore, we take into account only the leading order 
of $\om_i(\al,\th)$ in the small parameter $1/\gamma$. 
The first four coefficients of 
Eq.~(\ref{R_eta_2}) are then given by the Eqs.~(\ref{om_0123}), together 
with Eq.~(\ref{om_rel_end}). 
\begin{table}[hbt]
\vspace{0.1cm}
\begin{center}
\begin{tabular}{|c|@{\hspace{0.5cm}}l@{\hspace{0.5cm}}
                |c||@{\hspace{0.5cm}}l@{\hspace{0.5cm}}|}\hline 
 & & & \\[-0.2cm]
  $\th = \dsty\frac{\kB\Telec}{m_{\rm e}} $
        & $\;\tau_0$ & $\;\;M\;\;$ & $\;\;\al$ \\[0.3cm] \hline \hline
 50/511 & 0.05     & 0   & 2.83 \\ \hline
 50/511 & 0.05     & 1   & 2.80 \\ \hline
 50/511 & 0.05     & 2   & 2.75 \\ \hline
 50/511 & 0.05     & 3   & 2.74 \\ \hline
 50/511 & 0.05     & 4   & 2.74 \\ \hline \hline

 50/511 & 1.0      & 0   & 0.678 \\ \hline
 50/511 & 1.0      & 1   & 0.656 \\ \hline
 50/511 & 1.0      & 2   & 0.652 \\ \hline
 50/511 & 1.0      & 3   & 0.652 \\ \hline \hline

 50/511 & 3.0      & 0   & 0.186 \\ \hline
 50/511 & 3.0      & 1   & 0.179 \\ \hline
 50/511 & 3.0      & 2   & 0.179 \\ \hline \hline

  4.0   & 0.05     & 0   & 0.366 \\ \hline
  4.0   & 0.05     & 1   & 0.361 \\ \hline
  4.0   & 0.05     & 2   & 0.359 \\ \hline
  4.0   & 0.05     & 3   & 0.359 \\ \hline \hline

  4.0   & 1.0      & 0   & 0.0538 \\ \hline
  4.0   & 1.0      & 1   & 0.0480 \\ \hline
  4.0   & 1.0      & 2   & 0.0480 \\ \hline \hline

  4.0   & 3.0      & 0   & 0.0125 \\ \hline
  4.0   & 3.0      & 1   & 0.0103 \\ \hline
  4.0   & 3.0      & 2   & 0.0103 \\ \hline
\end{tabular}
\end{center}
\caption{Spectral index $\al$ for two plasma temperatures and different
         expansion parameters $M$ of the source function $B(\mu,\tau)$,
         and for Thomson optical half thickness $\tau_0 =0.05$, 1.0 
         and 3.0\,.}
\label{al_tab}
\end{table}
The spectral index
is not sensitive to the expansion of the $\mu$ and $\tau$ dependence
of $J(\mu,\tau)$. It is more than sufficient to choose $N=10$ in 
Eq.~(\ref{J_expand}), and $K=10$ in Eq.~(\ref{Q_expand}).
Results are given for an electron plasma temperature
 $\kB\Telec = \th m_{\rm e}$ of $5\,{\rm keV}$ and 
 $50\,{\rm keV}$ in the non-relativistic regime. In the relativistic 
regime we choose
 $\th = 4.0$ and
 $\th = 10.0$, corresponding roughly to $2$ and $5\,{\rm MeV}$ 
respectively.  
Figure \ref{al_tau_plot} shows the spectral
index $\al$ versus the Thomson optical half thickness $\tau_0$, using the 
expansion parameters given above ($L$, $M$, $N$ and $K$).

These values of $\al$ are in good agreement with those given by
Titarchuk \& Lyubarskij~(\cite{TitarLyu95}). These authors assumed 
an isotropic source function $B(\mu,\tau)$ (Eq.~\ref{source}),
which is certainly a good approximation for $\tau_0\gg 1$.
To relax this restriction, at least the first three expansion coefficients
of the source function must be taken into account ($M=0,1,2$), because
of the intrinsic \mbox{$1+(\eta')^2$} dependence of the Thomson scattering kernel.
Especially when the intensity $J(\mu,\tau)$ is highly anisotropic (which is
the case for $\tau_0\ll 1$, as shown in the next section) the source 
function $B(\mu,\tau)$ depends on the higher expansion coefficients
of the phase function, which leads to an anisotropy of $B(\mu,\tau)$.
In fact for $\tau_0 = 0.05$ the anisotropy
of the source function at the disk surface for $\th=50/511$ becomes
\eqb \label{B_anisotropy}
\frac{B(\mu =0,\tau_0 =0.05)}{B(\mu =1,\tau_0 =0.05)} = 2.3 \quad .
\eqe
The sensitivity of $\al$ to the angular expansion of the source function
is, however, not very strong, and it is well approximated 
by taking into account
the first four Legendre polynomials of the expansion 
($M=3$ in Eq.~(\ref{taylor})). This is shown for $\tau_0=0.05$, 1.0 and 3.0,
and two different plasma temperatures in Table \ref{al_tab}. 
Note that even for 
a value of $\al =2.74$ it is necessary to choose a Taylor expansion of
sufficiently high order (here: $L=16$), for {\em any} value of $M$.

Titarchuk \& Lyubarskij (\cite{TitarLyu95}) suggested interpolation formulas
(see Eq.~(17) and Eq.~(21) therein, and also Eq.~(27) in 
Titarchuk~\cite{Titar94}), which are valid for all optical depths and all 
plasma temperatures. 
The values of the spectral index given by these formulas, differ by
less then $1\%$ for
 $\tau_0\gg 1$ from those given in Fig.~\ref{al_tau_plot}; the largest 
discrepancy is $10\% $ at smaller values of $\tau$ (see also
Table~\ref{al_tab}).
\subsection{Angular distribution}
\label{angular_distr}
\begin{figure}[t]
    \vspace{-2.5cm}
    \begin{center} 
       \epsfxsize9.0cm 
       \mbox{\epsffile{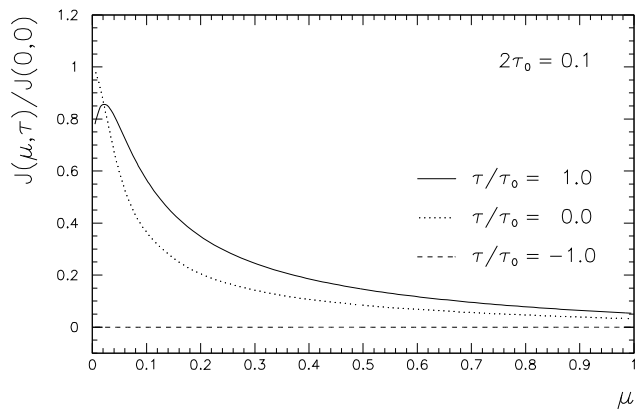}}
    \end{center}
    \vspace{-1cm} 
    \caption{Angular dependence of the specific intensity $J(\mu,\tau)$ for
             a Thomson optical thickness of $2\tau_0=0.1$.
             The intensity at the surface of the disk is given by
             $\tau/\tau_0 = 1.0$. The anisotropy of the integrated
             intensity is 
             $A(\tau_0)\equiv(\Jpar-\Jperp)/(\Jpar+\Jperp)=0.82\pm 0.03$
             for a wide temperature range. (Expansion up to $N=16$ and
             $K=10$). }
    \label{j_mu_1_plot}
\end{figure}
As described at the end of Sect.~\ref{method}, a singular value
decomposition of the matrix $\matr{\tens{F}}$ gives $J(\mu,\tau)$ in 
the form
of a polynomial of order $K$ in $\tau/\tau_0$, and of order $N$ in $\mu$
(for any set of parameters $\th$, $\tau_0$ and $\al$).
We choose $L=16$ and $M=6$ (see Eq.~(\ref{taylor})) for the phase function 
representation, and $\th=5/511$ for 
the plots discussed below. The minimum expansion for the angular and
spatial dependence ($K$ and $N$) have to be chosen differently for 
each optical depth (see figure captions). 
\begin{figure}[t]
    \vspace{-2.8cm}
    \begin{center} 
       \epsfxsize9.0cm 
       \mbox{\epsffile{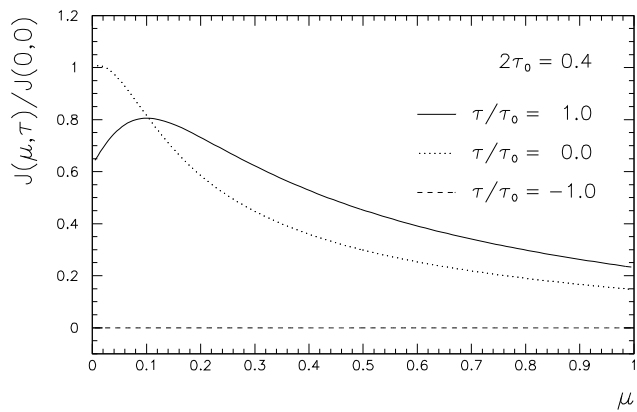}}
    \end{center}
    \vspace{-1cm} 
    \caption{Angular dependence of the specific intensity $J(\mu,\tau)$ for
             a Thomson optical thickness of $2\tau_0 = 0.4$.
             The intensity at the surface of the disk is given by
             $\tau/\tau_0 = 1.0$. The anisotropy of the integrated
             intensity is 
             $A(\tau_0)\equiv(\Jpar-\Jperp)/(\Jpar+\Jperp)=0.47\pm 0.03$
             for a wide temperature range. (Expansion up to $N=12$ and
             $K=14$).}
    \label{j_mu_2_plot}
\end{figure}
\begin{figure}[ht]
    \vspace{-3.0cm}
    \begin{center} 
       \epsfxsize9.0cm 
       \mbox{\epsffile{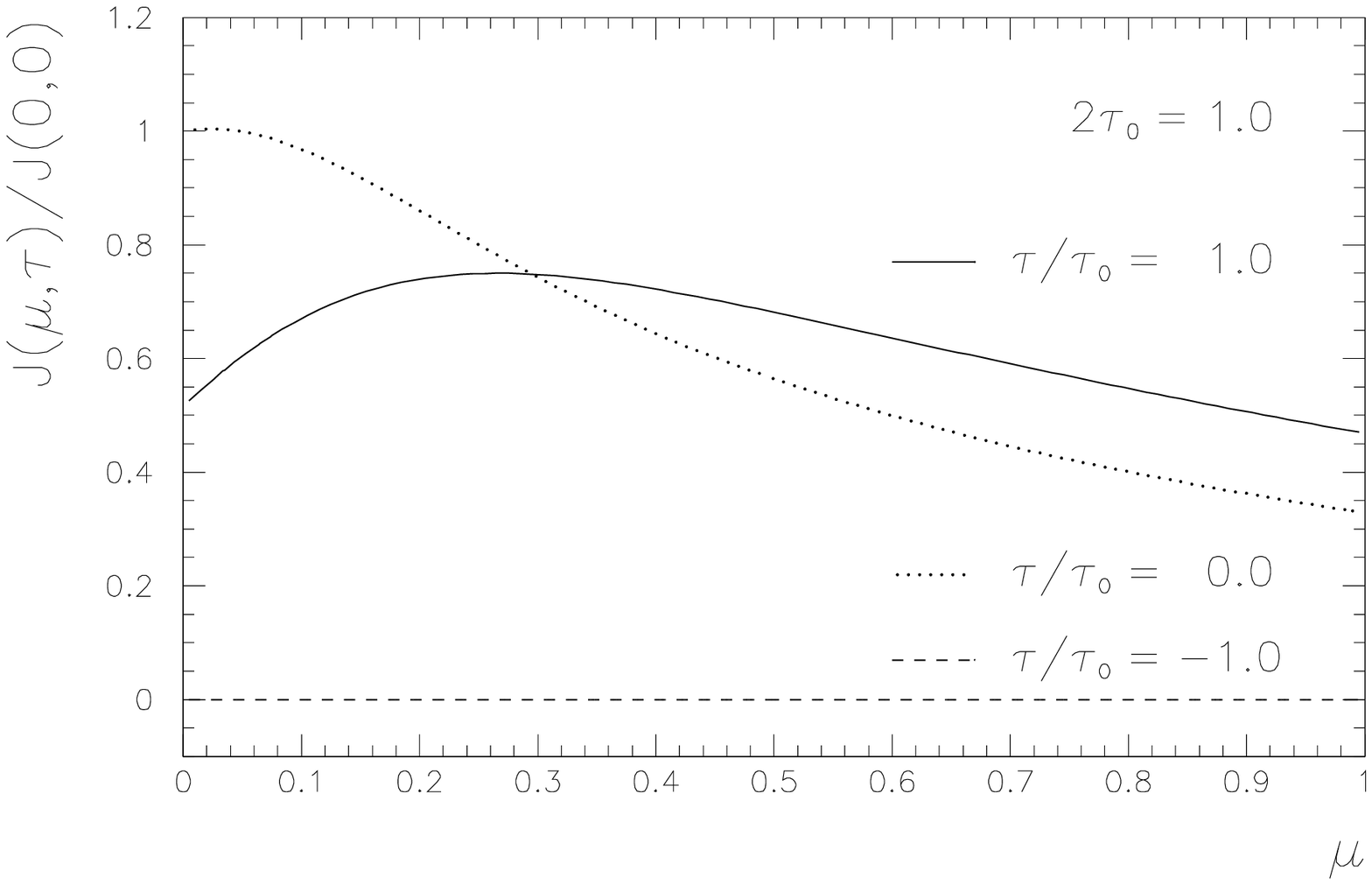}}
    \end{center}
    \vspace{-1cm} 
    \caption{Angular dependence of the specific intensity $J(\mu,\tau)$ for
             a Thomson optical thickness of $2\tau_0=1.0$.
             The intensity at the surface of the disk is given by
             $\tau/\tau_0 = 1.0$. The anisotropy of the integrated
             intensity is 
             $A(\tau_0)\equiv(\Jpar-\Jperp)/(\Jpar+\Jperp)=0.08\pm 0.02$
             for a wide temperature range. (Expansion up to $N=8$ and
             $K=16$).}
    \label{j_mu_3_plot}
\end{figure}

Because we are interested especially in {\em non}-isotropic intensities,
we define a measure of the anisotropy of $J(\mu,\tau)$. First, let
 $\Jpar(\tau)$ and $\Jperp(\tau)$ be the integrated intensity parallel
and perpendicular to the disk, over an interval of 
 $\Delta\mu = 0.1$:
\eqb \label{Jperp_Jpar}
\Jpar(\tau)  &:=& \int\limits_{0}^{0.1}\md\mu \, J(\mu,\tau) \,\,,\nn\\
\Jperp(\tau) &:=& \int\limits_{0.9}^{1}\md\mu \, J(\mu,\tau) \,\,.
\eqe
 From this we define an asymmetry\footnote{In analogy to the forward-backward 
asymmetry of the electro-weak interaction.} $A$, which is 0 for isotropic 
intensity, and 1 for extremely focussed intensity along the disk surface:
\eqb \label{anisotropy}
A(\tau_0) := \frac{\Jpar(\tau_0) - \Jperp(\tau_0)}
                  {\Jpar(\tau_0) + \Jperp(\tau_0)}\,.
\eqe
At an optical thickness of order unity ($2\tau_0\simeq 1$), this anisotropy
is about zero (see Fig.~\ref{j_mu_3_plot}). For smaller values of
 $\tau_0$, $A(\tau_0)$ can become very 
close to 1, as shown in Fig.~\ref{j_mu_1_plot}, where $J(\mu,\tau)$ is plotted
for an optical depth of $2\tau_0 = 0.1$,
normalised to the intensity in the middle of the disk, parallel to the surface.
The boundary condition gives $J(\mu,\tau=-\tau_0)\equiv 0$ for 
\mbox{$0<\mu\le1$} (no radiation enters the disk from outside, see dashed line).
The dotted line shows the intensity in the middle of the disk 
($\tau =0$), whereas the solid line shows the intensity at the 
surface, given by $J(\mu,\tau=\tau_0)$.
\begin{figure}[t]
    \vspace{-3.0cm}
    \begin{center} 
       \epsfxsize9.0cm 
       \mbox{\epsffile{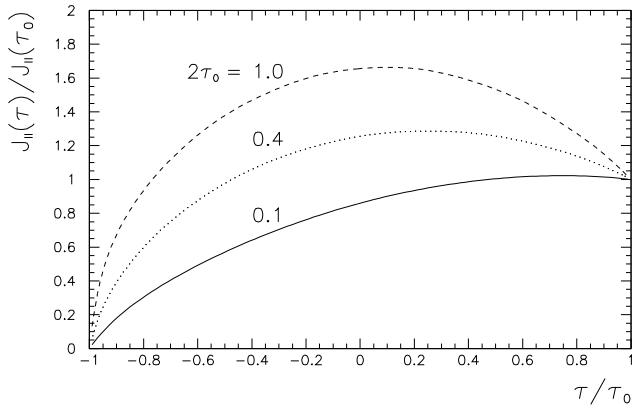}}
    \end{center}
    \vspace{-1cm} 
    \caption{Spatial distribution of the intensity parallel to the disk
             (as defined in Eq.~(\protect\ref{Jperp_Jpar})),
             normalised to the parallel intensity at the disk surface.
             Note that this plot shows the contributions of the half-space
             $0\le\mu\le1$.}
    \label{j_par_plot}
\end{figure}
The normalised specific intensity of the radiation is approximately
independent of the plasma temperature. As shown by Titarchuk \& Lyubarskij
(\cite{TitarLyu95}) the electron energy and photon spatial variables are 
completely decoupled if the source function is exactly isotropic. 
However, as discussed in the preceding section, the source function has a 
weak angular dependence, which leads to a coupling of the electron energy 
and photon spatial variables. This, in turn, 
implies that the angular dependence is a function of the plasma 
temperature. The range of variation over the temperature range 
considered is expressed
in form of an error of the anisotropy. The upper bound of the anisotropy is 
valid for $\th=5/511$, whereas the lower bound was calculated for $\th=100$.

Even for a moderately small optical depth of $2\tau_0 = 0.1$ 
(see Fig.\ref{j_mu_1_plot}) the anisotropy is $A(\tau_0)=0.82\pm 0.03$, 
which means that the radiation in the interval $0\le \mu\le 0.1$ parallel to
the disk surface, is a factor of about 10 more intense than the radiation 
in the interval $0.9 \le\mu\le 1$ perpendicular to the disk.
At increased angular resolution (smaller $\Delta\mu$) this factor 
is even larger.

The reason for the high anisotropy at small optical depth is the 
following:
photons which contribute to the power-law part of the spectrum
have to undergo a number of scatterings on electrons to
gain the required energy. The energy gain has a maximum for back-scattering
of the photons ($\Delta\theta =180^{\circ}$). Thus, those photons most 
effectively boosted in energy and least likely to escape the disk are 
those which move almost parallel to the surface.
This leads to a strong collimation in the disk plane for an optically 
thin disk. 
\subsection{Spatial distribution}
\label{spatial_distr}
The intensity $J(\mu,\tau)$ provides of course not only the angular
distribution, but also the spatial distribution. The solution discussed above
gives $J(\mu,\tau)$ for the half space $0\le\mu\le 1$. A full angle integrated
intensity at any space point is then the sum of the two half space intensities,
which are symmetric with respect to the middle of the disk. To compare the
spatial distribution to the previous figures, we show $J(\mu,\tau)$ also
for the half space $0\le\mu\le 1$, and used the same set of parameters
as in Sect.~\ref{angular_distr}. 

\begin{figure}[t]
    \vspace{-3.0cm}
    \begin{center} 
       \epsfxsize9.0cm 
       \mbox{\epsffile{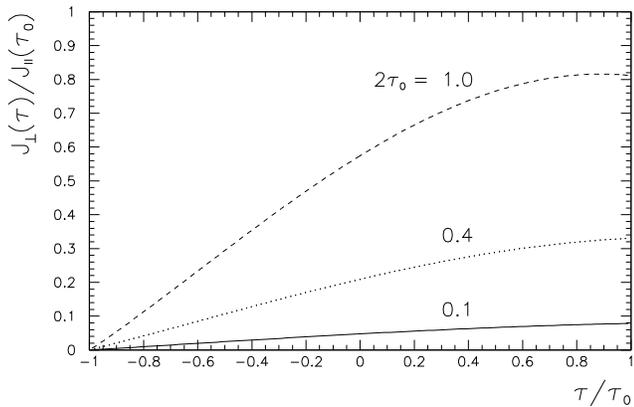}}
    \end{center}
    \vspace{-1cm} 
    \caption{Spatial distribution of the intensity perpendicular to the disk
             (as defined in Eq.~(\protect\ref{Jperp_Jpar})),
             normalised to the parallel intensity at the disk surface.
             Note that this plot shows the contributions of the half-space
             $0\le\mu\le1$.}
    \label{j_perp_plot}
\end{figure}

Figure \ref{j_par_plot} shows the intensity $\Jpar$ parallel to the disk, as 
defined in Eq.~(\ref{Jperp_Jpar}) for 
various optical depths, normalised to the parallel intensity at the disk 
surface. At $\tau=-\tau_0$ the intensity has to be 0, because we take
into account only the half space $0\le\mu\le 1$, for which the boundary
condition is $J(\mu,\tau=-\tau_0)\equiv 0$.
Figure \ref{j_perp_plot} shows the intensity $\Jperp$ perpendicular to the
disk, normalised to the parallel intensity $\Jpar$ at the disk surface. 
Note the difference in scale between this plot and Fig.~\ref{j_par_plot}. 
A comparison of Fig.~\ref{j_par_plot} and Fig.~\ref{j_perp_plot} provides 
the anisotropy $A(\tau)$ for all values of $\tau$. 
Because the perpendicular intensity $\Jperp$ drops faster for smaller values 
of $\tau/\tau_0$ than the parallel intensity
 $\Jpar$, the anisotropy $A(\tau < \tau_0)$ is even 
larger than $A(\tau_0)$, which is given in the captions of the 
Figs.~\ref{j_mu_1_plot} -- \ref{j_mu_3_plot}. 
\section{Discussion}
\label{discussion}
In this paper we have presented a new, semi-analytic method of 
obtaining solutions to the comptonisation problem, and used it to find 
the power-law index of photons scattered in the Thomson regime, 
neglecting the recoil of the scattered electron. Because these photons 
undergo a large number of scatterings before emerging from the plasma, 
they have \lq forgotten\rq\ the details (spatial and angular 
dependences) of their injection at low frequency. Once in the power-law 
region, which exists even for optically thin plasmas, the spatial and angular
dependence of the specific intensity of 
radiation ceases to be a function of frequency -- it is determined by 
a particular eigenfunction of the reduced transfer equation and depends only 
on the optical depth and temperature of the plasma.  
We compute this eigenfunction.
This distinguishes our approach from other numerical techniques in
the literature. Haardt~(\cite{haardt93}), for example, uses an approximation 
method in which anisotropy is accounted for only in the first scattering 
undergone by a photon, so that his results are accurate only fairly close to
the frequency of injection. Poutanen \& Svensson~(\cite{poutanensvensson96})
have developed a comprehensive code based on the iterative scattering method,
which treats the anisotropy \lq exactly\rq\ (on a discrete grid) and 
accounts for several processes which we neglect. This method converges 
rapidly provided only a few photon scatterings are important, i.e., at 
small optical depth and high temperature. However, it would need a large
number of grid points in order to resolve the sharp dependence of
the radiation intensity on angle such as displayed in Fig.~\ref{j_mu_1_plot}.
In principle, it 
should be possible to extend our technique to 
solve \lq 
inhomogeneous\rq\ problems, where the emergent spectrum depends on the 
input radiation. However, in this case it would be necessary to 
compute several eigenfunctions. Further extensions of the method to geometries 
other than slab, or to arbitrary electron distributions are straightforward.

To apply our results to observations of astrophysical objects we have
to consider the range of spectral indices of order 3 or less, because higher 
values
are probably too steep to be observable. A spectral index in this range can be 
achieved by a non-relativistic plasma with optical thickness of 
 $\tau_0\ga 1\,$, or by a relativistic plasma with optical thickness of
 $\tau_0\la 1$.
The assumption of Thomson scattering (as opposed to Klein-Nishina) 
restricts the relevant frequency range to  $x<1/\langle\gamma\rangle$  (where
 $\langle\gamma\rangle$ is the averaged Lorentz factor according 
to Eq.~(\ref{int_f})), i.e. to X-rays for the highest temperatures considered 
here. We also require the dominant source 
of soft photon input to be at a frequency which is low enough to 
require more than a single scattering before X-ray frequencies are 
achieved, otherwise the pure power-law spectrum is not achieved. The 
average frequency change  on scattering is given by 
 $\langle\Delta x\rangle /x \simeq 4\th$ in the non-relativistic regime, and
 $\langle\Delta x\rangle /x \simeq (4\th)^2$ for 
relativistic temperatures (see Pozdnyakov et al.~\cite{Pozdny83}).

As an example, consider a plasma with temperature $\kB\Telec=50\,{\rm keV}$
and input photons with energy $5\,{\rm eV}$. After 20 scatterings, which
removes all information of their initial distribution, they achieve
an energy of roughly $4\,{\rm keV}$.
The resulting normalised photon energy is then $x=0.008$, which is well
below $1/\langle\gamma\rangle = 0.86$. At an optical depth of
 $2\tau_0=0.4$
the intensity at the disk surface, parallel to it ($\Jpar$) becomes
three times the intensity perpendicular to the surface ($\Jperp$),
($A(\tau_0)=0.49$, compare Fig.~\ref{j_mu_2_plot}). 
The spectral index for these values of temperature and optical depth
is $\al = 1.77$. For relativistic 
temperatures, where spectral indices between 0 and 1 can be produced 
in a thin plasma disk ($\tau_0\ll 1$), which leads to a very high
anisotropy $A(\tau_0)$, the input photon energy has to be very
much lower than $5\,{\rm eV}$, in order to have more than 1 scattering
which shifts the photon energy into the X-ray regime.

In particular, our results concerning the degree 
of anisotropy (which is only weakly dependent on temperature) are 
relevant to the case of Seyfert galaxies, where plasma temperatures of 
the order of $100\,{\rm keV}$ have  been suggested (Titarchuk \& 
Mastichiadis~\cite{titarchukmastichiadis94}; 
Zdziarski et al.~\cite{zdziarskietal95}). In these objects, the ratio of 
optical  to X-ray 
luminosity should be  much smaller for objects seen \lq edge-on\rq\ 
than  for those seen 
\lq face-on\rq\ (an effect predicted by Haardt \& 
Maraschi~\cite{haardtmaraschi93}). It may, however, be difficult to 
disentangle this effect from that of the increased absorption of 
optical radiation expected in edge-on sources.

Quite apart from application to astrophysically important objects, the 
results of our computations should be helpful as a check on other 
numerical methods of solution, in particular Monte-Carlo simulations. 
Here it is worth mentioning that we have used a polarisation averaged 
treatment of the transport. All current Monte-Carlo codes also 
use this same approximation, so that the results they obtain are  
directly comparable to ours. Generalisation to polarisation dependent transfer 
is in principle possible.
\begin{acknowledgements}
We would like to thank A.~Mastichiadis for useful discussions and L. Titarchuk
for a thorough reading of the manuscript and several suggested improvements.
\end{acknowledgements}
\appendix
\section{Reduction of the Boltzmann equation to an algebraic
eigenvalue equation}
\label{appendix}
To express the Boltzmann equation in the form of an algebraic eigenvalue
equation, we express the angular and spatial dependence of the
source function  $R(\eta)$ and the specific intensity $J(\mu,\tau)$ in
terms of Legendre and Chebyshev polynomials. 

Let us first write the source function as a series
\eqb \label{R_eta}
 R(\eta) = \sum_{i=0}^M \om_i(\al,\th) P_i(\eta) \,,
\eqe
where the angular dependence of $R(\eta)$ is expressed in form of Legendre 
polynomials, with the normalisation:
\eqb \label{leg_norm}
\int\limits_{-1}^1 P_i(\mu) P_j(\mu)\md\mu =\frac{2}{2i+1}\,\,\delta_{ij}\,.
\eqe
These polynomials obey an addition theorem (see e.g., 
Landau \& Lifschitz \cite{LanLif} Eq.~(c.10)):
\eqb \label{add_theo}
P_l(\eta) &=& P_l(\mu) P_l(\mu_1) \\ 
          & & + \sum_{m=1}^l 2 \frac{(l-m)!}{(l+m)!}
            P_l^m(\mu) P_l^m(\mu_1) \cos(m(\phi-\phi_1)) \,, \nn
\eqe
where $\phi,\, \mu \equiv \cos\theta$ and $\phi_1,\, \mu_1\equiv \cos\theta_1$ 
define two directions with $\gamma\equiv\arccos\eta$ the angle between
these directions.
Equation~(\ref{R_eta}) can then be written:
\eqb \label{R_eta_2}
\frac{1}{2\pi} \int R(\eta) \md\phi = \sum_{i=0}^M \om_i(\al,\th)
   P_i(\mu) P_i(\mu_1) =: K(\mu,\mu_1) \,. 
\eqe

To calculate the coefficients $\om_i(\al,\th)$ for given values of the
temperature $\th$, we have to perform the integral over the electron
velocity $v$, which is tractable in the limit of high or low electron plasma
temperature. In these limits we can expand the integrand into a series 
in $v$, as described in Sect.~\ref{theta_small} for the 
non-relativistic case, and in $1/\gamma$, as described in 
Sect.~\ref{theta_big} for the relativistic case. 

The expansion of the angular and spatial dependence of  $J(\mu, \tau)$ due 
to Legendre and Chebyshev polynomials is described in 
Sects.~\ref{angular_dep} and \ref{spatial_dep}.
\subsection{Expansion of the phase function in the limit $\th \ll 1$}
\label{theta_small}
For small plasma temperatures the main contribution to 
the integral over electron
velocity arises from the region of small velocities. Therefore, the
integrand (more precisely the factor multiplying 
$f(v)$) can be expanded into a
Taylor series at $v=0$ with expansion coefficients $a_i$:
\eqb \label{taylor}
R(\eta) &=& \frac{3}{4}\int\md^3v f(v)\frac{1}{\gamma^2}
      \frac{(1-\tilde{\mu}_1 v)^{\al+1} }{(1-\tilde{\mu} v)^{\al+2} }\nn\\
 && \qquad\cdot\Bigl\{ 1+\Bigl(1-\frac{1-\eta}{\gamma^2
                (1-\tilde{\mu}v)(1-\tilde{\mu}_1 v)}\Bigr)^2 \Bigr\}\nn\\
 &=&\!\!\int\! v^2 \md v\,\md\tilde{\mu}_1\,\md\tilde{\phi}_{\rm e}\, f(v)\,
     \sum_{i=0}^L a_i(\al,\tilde{\mu}(\tilde{\mu}_1,\tilde{\phi}_{\rm e},\eta),
                       \tilde{\mu}_1,\eta )\,v^i \nn \\
 &=&  \sum_{i=0}^M \om_i(\al,\th) P_i(\eta) \,.
\eqe
We choose the direction of the outgoing photon
as $z$-axis. The cosine of the polar angle of the electron is then
 $\tilde{\mu}_1$ and 
the azimuth of the electron direction is denoted by $\tilde{\phi}_{\rm e}$. 
With this choice of coordinate system, the integration over electron direction 
($\md\tilde{\mu}_1\,\md\tilde{\phi}_{\rm e}$) becomes trivial. 
The remaining electron velocity integral can be expressed
in the form of moments, which can easily be calculated  
numerically:
\eqb \label{int_f}
\langle v^k \rangle := 4 \pi \int\limits_0^1 f(v) v^{k+2} \, \md v  \,.
\eqe
The normalisation of $f(v)$ gives $\langle v^0 \rangle = 1$, and 
 $\langle v^2 \rangle = 3\th $
for non-relativistic electron plasma temperatures. We used {\em Mathematica}
(see e.g. Wolfram \cite{Wolfram91}) to expand the integrand of $R(\eta)$ up to 
16th.\ order in $v$. Therefore
 $\langle v^k \rangle$ has to be calculated for $ k= 0,2,4,\dots ,16$.
The odd moments vanish due to the angular integration. 
This expansion in $v$ results in a polynomial series of order  $M=L/2+2$
in $\eta$. (Note, however, that one may still choose 
to truncate at $M\le L/2+2$).
As an example we give the expansion up to 4th.\ order in $v$. The coefficients
are then given by:
\eqb \label{omega_nr}
\om_0(\al,\th) &=& 1 + \frac{\vzw}{3}\, (\al^2+3\al) \nn \\ 
 & & + \frac{\vvi}{150}\, (\al^2+3\al)(7\al^2+21\al+22)\,,\nn \\[0.2cm]
\om_1(\al,\th) &=& - \frac{2}{5}\, \vzw \, (\al^2+3\al+1) \nn \\
 & & -\frac{\vvi}{25} \, (2\al^4+12\al^3+21\al^2+9\al+6)\,,\nn \\[0.2cm]
\om_2(\al,\th) &=& \frac{1}{2} + \frac{\vzw}{6}\, (\al^2+3\al-6) \nn \\
 & & +\frac{\vvi}{210}\, (10\al^4+60\al^3+55\al^2-105\al+78)\,,\nn \\[0.2cm] 
\om_3(\al,\th) &=& - \frac{\vzw}{10}\, (\al-1)(\al+4) \nn \\
 & & -\frac{\vvi}{50}\, (\al-1)(\al+4)(\al^2+3\al-7)\,,\nn \\[0.2cm]
\om_4(\al,\th) &=& \frac{\vvi}{175}\, (\al-1)(\al-2)(\al+4)(\al+5)\,.
\eqe
For very small electron temperatures {\em and} sufficiently small values
of $\al$, the sum of Eq.~(\ref{taylor}) converges very quickly. 
If we truncate it at (e.g.) $L=2$, the coefficient of the isotropic part of 
Eq.~(\ref{R_eta_2}) is
\eqb
\om_0(\al,\th) = 1+\th\, (\al^2+3\al)\,,
\eqe
in agreement with Eq.~(18) of Titarchuk \& 
Lyubarskij (\cite{TitarLyu95})\footnote{ Equation~(A.16) should read: 
 $ \hat{C}_0=\dsty\frac{4}{3\gamma^2}\big\{ 1+\dsty\frac{v^2}{3}
\left[ \al(\al+3)+6 \right] \big\}\,. $ }.
\subsection{Expansion of the phase function in the limit $\th \gg 1$}
\label{theta_big}
In the relativistic limit, the expansion has to be done in a slightly 
different way, because the phase function contains singular parts at $v=1$.
Using the orthogonality of the Legendre polynomials, the coefficients 
of Eq.~(\ref{R_eta}) can be written:
\eqb \label{om_rel}
\om_i(\al,\th)=\frac{2i+1}{4\pi}\int R(\eta)P_i(\eta)\,\md\tilde\Omega \,,
\eqe
where $\md\tilde\Omega=\md\eta\,\md\tilde\phi$ with $\tilde\phi$ the azimuth
of the incoming photon with respect to the outgoing one, which defines the
 $z$-axis. In this reference frame, the electron volume-element can be written
 $ \md^3 v = v^2\md v\,\md\tilde\Omega_{\rm e}$. Using this relation, and the 
definition of the phase function, the expansion coefficients can be written as
\eqb \label{om_rel_2}
\om_i(\al,\th)= 3\pi \int\limits_0^1\md v \, v^2 \frac{f(v)}{\gamma^2}
     \, \hat\om_i(\al,\gamma)\,,     
\eqe
with the temperature independent kernel
\eqb \label{hat_om_def}
\hat\om_i(\al,\gamma) & = & \frac{2i+1}{(4\pi)^2} \int \left( \frac{D_1}{D}
                  \right)^{\al+2} \frac{1}{D_1} \nn \\
     & & \qquad\qquad \cdot \, \bigl\{1+(\eta')^2\bigr\} \, P_i(\eta)\, 
                  \md\tilde\Omega \, \md\tilde\Omega_{\rm e} \,.
\eqe
The axis of integration in this reference frame can be changed, so that 
the electron direction becomes the $z$-axis. This is expressed by 
 $ \md\tilde\Omega\, \md\tilde\Omega_{\rm e} = \md\tilde\Omega\, \md\tilde\Omega_1 $.
With this choice of coordinates, it is easy to perform a Lorentz boost
to the electron rest frame. The azimuths of the in- and outgoing photons do 
not change, and the angle between the photon direction and the boost 
direction changes according to
\eqb \label{mu_trans}
\tilde\mu = \frac{\tilde\mu' +v}{1+v \tilde\mu'} \,.
\eqe
The differential solid angle transforms as
\eqb \label{om_trans}
\md\tilde\Omega = \frac{\md\tilde\Omega'}{\gamma^2(1+v \tilde\mu')^2}\,.
\eqe
These transformations lead to
\eqb \label{om_rel_3}
\hat\om_i(\al,\gamma) &=& \frac{2i+1}{(4\pi)^2\gamma^2} \int \md\tilde\mu'\, 
          \md\tilde\mu_1'\, \md\tilde\phi\, \md\tilde\phi_1'  \nn \\
  & & \qquad \cdot \, \frac{( 1+v\tilde\mu')^{\al}}{(1+v\tilde\mu_1')^{\al+3}}
	  \bigl\{1+(\eta')^2\bigr\} \,  P_i(\eta) \,,
\eqe     
where the argument of the Legendre polynomials has to be inserted as
\eqb \label{eta_trans}
\eta =1-\frac{1}{\gamma^2}\frac{1-\eta'}{(1+v\tilde\mu')(1+v\tilde\mu_1')}\,.
\eqe
To solve the above integrals one might use integral tables, or computer
programs. We used {\em Mathematica} to find analytic solutions for 
 $ i=0,1,2,3 $ (see Titarchuk \& Lyubarskij \cite{TitarLyu95}, 
Eqs.~(A10)-(A15), 
for $i=0$). In the limit $v \rightarrow 1$ these solutions diverge.
Therefore we separated the leading order terms. Extracting 
a $\gamma$-dependent factor according to:
\eqb \label{gamma_extract}
\hat\om_i(\al,\gamma)=(2\gamma)^{2\al+2} \,\hat\om_i(\al) \,,
\eqe
these solutions are:
\eqb \label{om_0123}
\hat\om_0(\al) &=&  \frac{\al(\al+3)+4}
                          {(\al+1)(\al+2)^2(\al+3)} \,, \\[0.2cm]
\hat\om_1(\al) &=& -3 \, \frac{\al(\al+3)+4}
                          {(\al+2)^2(\al+3)^2} \,,\nn \\[0.2cm]
\hat\om_2(\al) &=& 5 \,\al \, \frac{\al(\al+3)+4}
                          {(\al+2)^2(\al+3)^2(\al+4)} \,,\nn \\[0.2cm]
\hat\om_3(\al) &=& 7 \,\al (1-\al)
                       \, \frac{\al(\al+3)+4}
                          {(\al+2)^2(\al+3)^2(\al+4)(\al+5)} \,.\nn
\eqe
Again, the isotropic part, $\hat\om_0(\al,\gamma)$, is in agreement with
Eq.~(A17) of Titarchuk \& Lyubarskij (\cite{TitarLyu95}).\\
For high plasma temperatures one can expand the integrand of 
Eq.~(\ref{om_rel_2})  into a series. Using the following relation:
\eqb \label{dv_dgamma}
v^2\,\md v = \frac{1}{\gamma^3} \sqrt{1-\frac{1}{\gamma^2}}\,\md \gamma = 
           \frac{1}{\gamma^3}(1-\frac{1}{2\gamma^2} + \dots) \md \gamma
\eqe
Eq.~(\ref{om_rel_2}) becomes:
\eqb \label{om_gamma_int}
\om_i(\al,\th) &=& \frac{3}{4}\, \frac{1}{\th \,{\rm K}_2(1/\th)}\,
                       \,\hat\om_i(\al)   \nn  \\ 
     & &  \,\, \cdot \int\limits_1^{\infty}  \md\gamma 
                 \Bigl(1-\frac{1}{2\gamma^2}\Bigr)
            (2\gamma)^{2\al+2}\, \mbox{exp}\bigl(-\frac{\gamma}{\th}\bigr)\,.
\eqe
The above integration can be expressed in terms of the incomplete
 $\Gamma$-function\footnote{$\Gamma(z,a):=\int_a^{\infty}t^{z-1}e^{-t}\md t $}.
Then the coefficients $\om_i(\al,\th)$ can be written:
\eqb \label{om_rel_end}
\om_i(\al,\th)&=& 3\frac{(2\th)^{2\al}}{{\rm K}_2(1/\th)}\,\hat\om_i(\al)\nn\\ 
     & &  \,\,\cdot \Bigl[ \th^2 \Gamma\bigl(2\al+3,\frac{1}{\th}\bigr)  
            - \frac{1}{2} \Gamma\bigl(2\al+1,\frac{1}{\th}\bigr)\Bigr]\,.
\eqe
Together with Eqs.~(\ref{om_0123}) this defines the 
expansion coefficients (for $i=0,1,2,3$) of the phase function for a high 
electron plasma temperature.  
\subsection{The angular dependence of $J(\mu, \tau)$}
\label{angular_dep}
The angular dependence of the intensity $J(\mu,\tau)$ can be expressed 
in the form
of a series of Legendre polynomials with $\tau$-dependent coefficients. 
Because of the symmetry of the boundary conditions, it is necessary only to 
calculate  $J(\mu,\tau)$ for one half-space: $0\le \mu \le 1$ and 
 $-\tau_0\le\tau\le\tau_0$. The intensity 
in the second half-space is then given by 
\eqb \label{symmetry}
 J(-\mu,\tau)=J(\mu,-\tau) \,.
\eqe
Having in mind the orthogonality relation for Legendre polynomials on the interval
(e.g.) $0\le \mu \le 1$ :
\eqb \label{ortho}
\int\limits_0^1 P_n(2\mu-1) P_k(2\mu-1)\md\mu =\frac{1}{2 n+1}\delta_{nk}\,,
\eqe
we expand the intensity as follows
\eqb \label{J_expand}
J(\mu,\tau)\big|_{\mu \le 0} &=& \sum_{n=0}^N \frac{2 n+1}{2} P_n(2\mu+1)
                                        Q_n^-(\tau) \nn \,, \\
J(\mu,\tau)\big|_{\mu \ge 0} &=& \sum_{n=0}^N \frac{2 n+1}{2} P_n(2\mu-1)
                                        Q_n^+(\tau)  \,.
\eqe
Defining the new variables $\xi$ and $\zeta$, according to:
\eqb \label{def_x_y}
\zeta &:=& 2\mu+1 \big|_{\mu<0} \nn \,, \\
\xi &:=& 2\mu-1 \big|_{\mu>0} \,,
\eqe
the inversion of Eq.~(\ref{J_expand}) can be written as
\eqb \label{Q(J)}
Q_n^-(\tau) &=& \int\limits_{-1}^1 J\big(\frac{\zeta-1}{2},\tau\big) 
                                                 P_n(\zeta)\md \zeta\nn\,,\\
Q_n^+(\tau) &=& \int\limits_{-1}^1 J\big(\frac{\xi+1}{2},\tau\big) 
                                                 P_n(\xi) \md \xi  \,.
\eqe
Noting that $P_n(\mu)=(-1)^n P_n(-\mu)$, 
the symmetry condition (Eq.~\ref{symmetry}) gives
\eqb \label{Q+Q-}
Q_n^-(\tau) = (-1)^n Q_n^+(-\tau) \,.
\eqe
It is useful to rewrite the kernel $K(\mu,\mu_1)$ using a 
transformation of variables and to 
split it into two parts according to $\mu_1<0$ and 
 $\mu_1\ge 0$. This is done in the following way: 
written in the form of a scalar product, 
the right-hand side of Eq.~(\ref{R_eta_2}) reads:
\eqb \label{K_matrix}
K(\mu,\mu_1) = \vec{P}(\mu)\, \tens{\om}(\al,\th)\, \vec{P}(\mu_1) 
\eqe
with a diagonal, $M+1$--dimensional matrix $\tens{\om}(\al,\th)$ and vectors
 $\vec{P}$. Let $\tens{W}$ be the matrix, which transforms the Legendre 
polynomial 
as\footnote{
The matrix \protect\tens{W} can be calculated from Eq.~(A28) by expressing
the vectors $\vec{P}(\mu_1)$ and $\vec{P}(2\mu_1-1)$ in the common basis
$(1,\mu,\mu^2,\dots,\mu^M)$ and inverting the matrix of coefficients
of $\vec{P}(\mu_1)$.}
\eqb \label{P_trans}
\vec{P}(2\mu_1-1) = \tens{W}\,\, \vec{P}(\mu_1) \,.
\eqe
Then, for $\mu_1\ge 0$, one can write:
\eqb \label{def_om^+}
\tens{\om}(\al,\th)\,\vec{P}(\mu_1) &=& \tens{\om}(\al,\th)\,
            \tens{W}^{-1}\,\tens{W}\,\vec{P}(\mu_1) \nn \\ 
  &=:&  \tens{\om}^+(\al,\th)\,\vec{P}(2\mu_1-1) \,, 
\eqe
with an $M+1$--dimensional, {\em non}-diagonal, matrix $\tens{\om}^+(\al,\th)$.
An analogous transformation in the range $\mu_1<0$ leads to
\eqb \label{K_trans}
K(\mu,\mu_1) &=& \vec{P}(\mu)\,\Big[\,\, \tens{\om}^+(\al,\th)\, 
                         \vec{P}(2\mu_1-1) \, \mbox{H}(\mu_1) \nn \\
    & & \qquad + \,\, \tens{\om}^-(\al,\th)\, \vec{P}(2\mu_1+1) \,
                                 \mbox{H}(-\mu_1)\,\Big] \,,
\eqe
with the Heaviside function:
\eqb \label{theta}
\mbox{H}(\mu_1) = \left\{\,\, \begin{array}{r@{\quad:\quad}l}
                    1 & \mu_1 \ge 0 \\ 0 & \mu_1<0 \quad . 
                              \end{array} \right.  
\eqe
The source function $B(\mu,\tau)$ (Eq.~\ref{source}) becomes with this 
(and the use of Eq.~(\ref{Q(J)}),
and suppressing the dependence of $\tens{\om}^{\pm}$ on $\al$ and $\th$):
\eqb \label{B_trans}
B(\mu,\tau) = \frac{1}{4}\bigl[\,\vec{P}(\mu)\,\tens{\om}^+ \,\vec{Q}^+(\tau)
                             + \,\vec{P}(\mu)\,\tens{\om}^- \,\vec{Q}^-(\tau)
                                        \bigr] \,.
\eqe
Inserting this into the Boltzmann equation (Eq.~\ref{Boltzmann}), 
substituting $\xi=2\mu-1$ for
 $\mu \ge 0$, multiplying by $P_k(\xi)$, and integrating 
$\xi$ from -1 to 1, we obtain
\eqb \label{Boltzmann_integral}
\lefteqn{
\frac{\partial}{\partial\tau}\int\limits_{-1}^1\md\xi\,\frac{\xi+1}{2} P_k(\xi)
       J\bigl(\frac{\xi+1}{2},\tau \bigr) = } \nn \\ 
 &&\,\,\frac{1}{4}\sum_{i,j=0}^N\int\limits_{-1}^1\md \xi \,\,P_k(\xi)
                                       P_i\bigl(\frac{\xi+1}{2}\bigr)
   \,\bigl[\,\om_{ij}^+\,Q_j^+(\tau) +\om_{ij}^-\,Q_j^-(\tau)\bigr] \nn \\
 &&\,\,-\int\limits_{-1}^1 \md\xi\, P_k(\xi) \, 
                              J\bigl(\frac{\xi+1}{2},\tau\bigr) \,.
\eqe
The indices $i,j,k$ run from 0 to $N$. If $N>M$ the matrices
 $\om_{ij}^+$ and $\om_{ij}^-$ can be defined with all elements equal to 0
for $i,j>M$. If $N<M$ only the elements with $i,j \le N$ are taken into 
account, so that in both cases the matrices formally become $N+1$--dimensional.
Using the recurrence relation for Legendre polynomials, 
\eqb \label{recurrence}
\xi\,P_k(\xi) = \frac{k}{2k+1}\,P_{k-1}(\xi) +\frac{k+1}{2k+1}\,P_{k+1}(\xi)\,,
\eqe
the left hand side of Eq.~(\ref{Boltzmann_integral}) can be written in terms of
a tridiagonal matrix, whose elements are defined as:
\eqb \label{X_def}
Z_{kl} := \frac{k}{2k+1} \, \delta_{k,l+1} + \delta_{kl} 
                        +\frac{k+1}{2k+1} \, \delta_{k,l-1} \,.
\eqe
We further define a matrix with the elements:
\eqb \label{S_def}
S_{ki} :=\frac{1}{2}\int\limits_{-1}^1\md\xi\,P_k(\xi)
                                         P_i\bigl(\frac{\xi+1}{2}\bigr)\,.
\eqe
Using these definitions, and again Eq.~(\ref{Q(J)}),
Eq.~(\ref{Boltzmann_integral}) becomes
\eqb \label{Boltzmann_2}
\sum_{l=0}^N Z_{kl} \frac{\partial Q_l^+(\tau)}{\partial \tau} & = & 
 \sum_{i,j=0}^N S_{ki}\,\bigl[\,\om_{ij}^+ Q_j^+(\tau) + 
                                \om_{ij}^- Q_j^-(\tau) \bigr] \nn\\ 
   & & -  2\,Q_k^+(\tau) \,.
\eqe
Having in mind the symmetry relation, Eq.~(\ref{Q+Q-}), and defining a diagonal
matrix according to
\eqb \label{E_def}
 E_{jm} := (-1)^{j} \,\delta_{jm} \,,
\eqe
Eq.~(\ref{Boltzmann_2}) reads in matrix form
\eqb \label{Boltzmann_matrix}
\tens{Z}\,\frac{\partial \vec{Q}^+(\tau)}{\partial \tau} &=&
     \tens{S}\,\bigl[\,\tens{\om}^+\,\vec{Q}^+(\tau)\,+\,
                \tens{\om}^-\,\tens{E}\,\vec{Q}^+(-\tau)\,\bigr] \nn\\
 & & -2\, \vec{Q}^+(\tau) \,.
\eqe
This can be expressed in somewhat shorter way, using the definitions
\eqb \label{M_def}
\tens{M} &:=& -2\, \bbbone + \tens{S}\,\tens{\om}^+ \,,\nn\\[0.2cm]
\tens{\tilde{M}} &:=&  \tens{S}\,\tens{\om}^-\,\tens{E} \,.
\eqe
Now the Boltzmann equation can be written as
\eqb \label{Boltzmann_short}
\tens{Z}\,\frac{\partial \vec{Q}^+(\tau)}{\partial \tau} &=&
    \tens{M}\,\vec{Q}^+(\tau) + \tens{\tilde{M}}\,\vec{Q}^+(-\tau)\,,
\eqe
and represents a system of $N+1$ coupled differential equations.
 
For a given temperature in the non-relativistic regime \mbox{$(\th\ll 1)$,} 
the moments $\langle v^k\rangle$ are to be calculated
up to $k=L$. The matrix elements of $\tens{M}$ and $\tens{\tilde{M}}$
are then polynomials in $\al$ of maximal order $L$.

In case of high plasma temperature $(\th\gg 1)$, the matrix elements 
of $\tens{M}$ and $\tilde{\tens{M}}$ 
depend on the incomplete $\Gamma$-function (see Eq.~(\ref{om_rel_end})). 
\subsection{The spatial dependence of $J(\mu, \tau)$}
\label{spatial_dep}
Equation (\ref{Boltzmann_short}) is a system of coupled differential equations
for the $N+1$ expansion coefficients of $J(\mu,\tau)$, which contain the
$\tau$-dependence of the intensity. In expanding these coefficients 
themselves
into a
series, the problem is reduced to an algebraic eigenvalue problem.
A difference arises compared to the above treatment of the angular dependence 
in that we have to
take into account the boundary condition, which is that 
no radiation enter the
disk from outside:
\eqb \label{J_Randbed}
J(\mu,\tau_0)\big|_{\mu<0}\,\equiv\,0\,\equiv\,J(\mu,-\tau_0)\big|_{\mu>0}\,.
\eqe  
We introduce at this point a new variable according to
\eqb \label{tau->y}
y(\tau)=\frac{\tau}{\tau_0}\,.
\eqe
Using this transformation and Eq.~(\ref{J_expand}), the boundary condition 
reads for all expansion coefficients:
\eqb \label{Q_Randbed}
\vec{Q}^+(-1) = 0 = \vec{Q}^-(1)\,.
\eqe
We chose an expansion of $\vec{Q}^+$ in Chebyshev polynomials according 
to
\eqb \label{Q_expand}
\vec{Q}^+(y) = \vec{q}_0 + \sum_{i=1}^K \bigl[ 1-T_i(-y)\bigr]\,\vec{q}_i\,.
\eqe
with $\vec{Q}^-(y)$ given by the symmetry condition Eq.~(\ref{Q+Q-}).
The boundary condition then gives $\vec{q}_0 = 0$. Inserting this into 
Eq.~(\ref{Boltzmann_short}) one gets
\eqb \label{T_eq_1}
\frac{1}{\tau_0}\sum_{i=1}^K(-1)^{i+1}\frac{\partial T_i(y)}{\partial y}
  \tens{Z}\,\vec{q}_i &=&  \sum_{i=1}^K T_i(y) \bigl[(-1)^{i+1}\tens{M}-
                       \tens{\tilde{M}}\bigr]\,\vec{q}_i \nn\\ 
 & & + \sum_{i=1}^K \bigl[\tens{M}+\tens{\tilde{M}}\bigr]\,\vec{q}_i\,. 
\eqe
Multiplying this equation by $T_j(y)/\sqy$ and integrating $y$ from
$ -1$ to $1$, it becomes
\eqb \label{T_eq_2}
\lefteqn{
 \frac{1}{\tau_0}\sum_{i=1}^K(-1)^{i+1}\int\limits_{-1}^1\md y \,
        \frac{T_j(y)\frac{\partial}{\partial y}T_i(y)}{\sqy}
                         \tens{Z}\,\vec{q}_i =} \nn \\
 && \quad\qquad\sum_{i=1}^K \int\limits_{-1}^1\md y\,
                                    \frac{T_j(y)\,T_i(y)}{\sqy}
       \bigl[(-1)^{i+1}\tens{M}-\tens{\tilde{M}}\bigr]\,\vec{q}_i \nn \\
 && \qquad +\,\sum_{i=1}^K \int\limits_{-1}^1\md y\,\frac{T_j(y)}{\sqy}
       \bigl[\tens{M}+\tens{\tilde{M}}\bigr]\,\vec{q}_i \,.
\eqe   
This is a vector equation with $K\times K$ matrices, defined by the integrals
over Chebyshev polynomials. We denote these matrices by
\eqb \label{T_matr_def}
(\tens{T}_{\rm d})_{ji} &:=& \int\limits_{-1}^1 \md y\,\frac{T_j(y)
                  \frac{\partial}{\partial y} T_i(y)}{\sqy} \,,\nn\\
(\tens{T}_{\rm o})_{ji} &:=&\int\limits_{-1}^1\md y\,\frac{T_j(y) T_i(y)}{\sqy}
                                                                   \,,\nn\\
(\tens{T}_{\rm h})_{ji} &:=& \int\limits_{-1}^1 \md y\,\frac{T_j(y)}{\sqy} 
                                                                    \,,\nn\\
(\tens{E}_{\rm t})_{ik} &:=& (-1)^{i+1} \, \delta_{ik}\,.
\eqe
The matrix elements are straightforwardly calculated 
using the orthogonality relation
for Chebyshev polynomials:
\eqb \label{T_ortho}
\int\limits_{-1}^1 \md y\,\frac{T_j(y) T_i(y)}{\sqy} = 
         \left\{ \begin{array}{r@{\quad :\quad}l}
       0 & i\neq j \\ \pi/2 & i=j\neq 0\\ \pi & i=j=0 \quad,
                 \end{array} \right.
\eqe
the expression for the derivative:
\eqb
{\md T_i(y)\over \md y}={-i\,y\,T_i(y) + i\,T_{i-1}(y)\over (1-y^2)}\,,
\eqe
and the recurrence relation
\eqb
T_{i+1}(y)=2\,y\,T_i(y) - T_{i-1}(y)\,.
\eqe

The boundary condition $\vec{q}_0=0$ is not explicitly included 
in the above matrix equation, but must be inserted 
\lq by hand\rq\ into Eq.~(\ref{T_eq_2}). 
This is done by
extending the matrix equation to $i,j=0,\dots,K$, and adding the equations
\eqb \label{boundary_eqn}
\sum_{i=0}^K C_{ji}\, \vec{q}_i = 0
\eqe
with
\eqb \label{C_def}
(\tens{C})_{ji} =  \left\{ \begin{array}{r@{\quad :\quad}l}
       1 & j=K \quad\mbox{and}\quad i=0 \\ 0 & \mbox{otherwise} \quad.
                 \end{array} \right. 
\eqe
All other matrices have to be set to 0 for $j=K$. The matrices are then given
by 
\eqb \label{T_def}
(\tens{T}_{\rm d})_{ji} &=&  \left\{ \begin{array}{r@{\quad :\quad}l}
         i\,\pi & i>j \quad\mbox{and}\quad i+j=\,\mbox{odd} \\
              0 & \mbox{otherwise} \quad , 
                        \end{array} \right. \nn \\[0.3cm]
(\tens{T}_{\rm o})_{ji} &=&  \left\{ \begin{array}{r@{\quad :\quad}l}
         0      & i\neq j \quad\mbox{or}\quad j = K \\
         \pi/2  & i=j\neq 0 \quad\mbox{and}\quad j\neq K \\
         \pi    & i=j=0 \quad ,
                        \end{array} \right. \nn \\[0.3cm]
(\tens{T}_{\rm h})_{ji} &=&  \left\{ \begin{array}{r@{\quad :\quad}l}
            \pi & j=0  \\
              0 & \mbox{otherwise} \quad .
                        \end{array} \right.
\eqe
It is useful to write Eq.~(\ref{T_eq_2}) in a different way. It is a vector
equation in which the elements are themselves vectors, multiplied by matrices.
This form of matrix multiplication is just the outer product of matrices,
denoted by $\otimes$. The $K+1$ vectors of dimension $N+1$ can be written
as a common vector in a $(K+1)\cdot (N+1)$-dimensional product-space:
\eqb \label{q_vec}
\ul{\vec{q}} := \left( \begin{array}{c}
                \vec{q}_0 \\ \vec{q}_1 \\ \vdots \\ \vec{q}_K 
                 \end{array} \right) \,\, .  
\eqe
Note that the vector $\ul{\vec{q}}$ does not have an index but includes
 {\em all} expansion
coefficients of the angular and spatial dependence of $J(\mu,\tau)$.
Equation (\ref{T_eq_2}) now reads
\eqb \label{T_eq_outer}
\bigl[\,\frac{1}{\tau_0}\,(\tens{E}_{\rm t}\,\tens{T}_{\rm d})\otimes
    \tens{Z}\,\bigl]\ul{\vec{q}} &=& 
     \bigl[\, \tens{C}\otimes\bbbone\,+\,\tens{T}_{\rm h}\otimes
    (\tens{M}+\tilde{\tens{M}}) \nn\\
 &&  +\,(\tens{E}_{\rm t}\,\tens{T}_{\rm o})\otimes\tens{M}\,
      -\,\tens{T}_{\rm o}\otimes\tilde{\tens{M}}\,\bigr]\,\ul{\vec{q}}\,.
\eqe
Let us denote the matrix on the left hand side, which does not depend on
any parameter, by 
\eqb \label{A_def}
\matr{A} := (\tens{E}_{\rm t}\,\tens{T}_{\rm d})\otimes\tens{Z}\,,
\eqe 
and the right hand matrix, which depends on $\al$ and $\th$ (due to
 $\tens{\om}^{\pm}(\al,\th)$ in $\tens{M}$ and $\tilde{\tens{M}}$) by
 $\matr{D}(\al,\th)$. 
Equation (\ref{T_eq_outer}) reads
with these definitions:
\eqb \label{T_eq_short}
\bigl[\,\frac{1}{\tau_0}\matr{A}\,-\,\matr{D}(\al,\th)\,
  \bigr]\,\ul{\vec{q}}\,\equiv\,\matr{F}(\al,\th,\tau_0)\,\ul{\vec{q}}\,=0\,. 
\eqe 
This defines an algebraic eigenvalue problem for all of the expansion 
coefficients $\ul{\vec{q}}$.


\begin{thebibliography}{}
%
\bibitem[1971]{Cooper71}
Cooper G., 1971, Phys. Rev. D 3, 2312 
%
\bibitem[1996]{ebisawaetal96}
Ebisawa K., Titarchuk L., Chakrabarti S.K.,
1996, Publ.\ Astron.\ Soc.\ Japan 48, 59
%
\bibitem[1993]{haardt93}
Haardt F., 1993, ApJ 413, 680
%
\bibitem[1993]{haardtmaraschi93}
Haardt F., Maraschi L., 1993, ApJ 413, 507
%
\bibitem[1994]{haardtetal94}
Haardt F., Maraschi L., Ghisellini G., 1994, ApJ 432, L95
%
\bibitem[1995]{HuaTitar95}
Hua X.-M., Titarchuk L.G., 1995, ApJ 449, 188  
%
\bibitem[1976]{Katz76}
Katz J.I., 1976, ApJ 206, 910
%
\bibitem[1988]{LanLif}
Landau L.D., Lifschitz E.M., 1988, {\em Quantenmechanik}, Akademie-Verlag,
     Berlin
%
\bibitem[1973]{Pomraning73}
Pomraning G.C., 1973, {\em Radiation Hydrodynamics}, Pergamon Press, Oxford
%
\bibitem[1996]{poutanensvensson96}
Poutanen J., Svensson R., 1996, ApJ 470, 249
%
\bibitem[1983]{Pozdny83}
Pozdnyakov L.A., Sobol I.M., Sunyaev R.A., 1983, 
    Ap.\ Space Phys.\ Rev.\ 2, 189
%
\bibitem[1988]{Prasad88}
Prasad M.K., Shestakov A.I., Kershaw D.S., Zimmerman G.B., 1988,
    J. Quant. Spectrosc. Rad. Transf. 40, 29
%
\bibitem[1986]{Press86}
Press W.H., Flannery B.P., Teukolsky S.A., Vetterling W.T., 1986,
   {\em Numerical Recipes}, Cambridge University Press
%
\bibitem[1976]{Shapiro76}
Shapiro S.L., Lightman A.P., Eardley D.M., 1976, ApJ 204, 187
%
\bibitem[1995a]{sternetal95a}
Stern B.E., Begelman M.C., Sikora M., Svensson R., 1995a, MNRAS 272, 291
%
\bibitem[1995b]{sternetal95b}
Stern B.E., Poutanen J., Svensson R., Sikora M., Begelman M.C., 1995b, 
ApJ 449, L13
%
\bibitem[1980]{SunTitar80}
Sunyaev R.A., Titarchuk L.G., 1980, A\&A 86, 121
%
\bibitem[1985]{SunTitar85}
Sunyaev R.A., Titarchuk L.G., 1985, A\&A 143, 374 
%
\bibitem[1979]{SunTruem79}
Sunyaev R.A., Tr\"umper J., 1979, Nature 279, 506
%
\bibitem[1994]{Titar94}
Titarchuk L.G., 1994, ApJ 434, 570
%
\bibitem[1995]{TitarLyu95}
Titarchuk L.G., Lyubarskij Y., 1995, ApJ 450, 876 
%
\bibitem[1994]{titarchukmastichiadis94}
Titarchuk L.G., Mastichiadis A., 1994, ApJ 433, L33
%
\bibitem[1991]{Wolfram91}
Wolfram S., 1991, {\em Mathematica}, Addison-Wesley, New-York
%
\bibitem[1986]{Zdzia86}
Zdziarski A.A., 1986, ApJ 303, 94
%
\bibitem[1995]{zdziarskietal95}
Zdziarski A.A., Johnson W.N., Done C., Smith D., McNaron-Brown K.,
1995, ApJ 438, L63
%
\end{thebibliography}
\end{document}